\documentstyle[twocolumn,aps,psfig]{revtex}

\begin{document}
\draft

\twocolumn[\hsize\textwidth\columnwidth\hsize\csname @twocolumnfalse\endcsname

\title{Critical exponent in the magnetization curve of quantum spin
chains
}

\author{T\^oru Sakai$^{1}$ and Minoru Takahashi$^{2}$}
\address{
$^{1}$Faculty of Science, Himeji Institute of Technology, Kamigori,
Ako-gun, Hyogo 678-12, Japan\\
$^{2}$Institute for Solid State Physics, University of Tokyo, Roppongi,
Minato-ku, Tokyo 106, Japan\\
}

\date{January 98}
\maketitle 

\begin{abstract}
The ground state magnetization curve around the critical magnetic field
$H_c$ of quantum spin chains with the spin gap is investigated. 
We propose a size scaling method 
to estimate the critical exponent $\delta$ 
defined as $m\sim |H-H_c|^{1/\delta}$ from finite cluster calculation. 
The applications of the method to the $S=1$ antiferromagnetic chain 
and $S=1/2$ bond alternating chain lead to a common conclusion 
$\delta =2$. 
The same result is derived 
for both edges of the magnetization plateau of the 
$S=3/2$ antiferromagnetic chain with the single ion anisotropy. 
\end{abstract}

\vskip2pc]
\narrowtext

%
%

The magnetization curve of quantum spin chains 
shows various nontrivial behaviors due to quantum effects.
In the spin-gap systems, where a finite energy gap exists in 
the spin excitation spectrum, 
the gap is controlled by applied magnetic field $H$ 
through the Zeeman term in the Hamiltonian. 
The typical examples are the antiferromagnetic chain with integer spin
called Haldane magnets\cite{haldane}, 
spin Peierls systems and spin ladders etc. 
In these systems 
a phase transition occurs at the critical field $H_c$ corresponding 
to the amplitude of the gap\cite{sakai1,sakai2,sakai3}; 
the system has the nonmagnetic ground state and a finite gap for $H<H_c$, 
while the magnetic ground state and no gap for $H>H_c$. 
The transition was observed in the magnetization measurements 
on some quasi-one-dimensional materials; 
for example, 
an $S=1$ antiferromagnet 
Ni(C$_2$H$_8$N$_2$)$_2$NO$_2$(ClO$_4$)\cite{nenp1,nenp2}, 
abbreviated NENP, 
and a spin-Peierls compound CuGeO$_3$.\cite{cuge1,cuge2} 

In our previous work\cite{sakai1} 
we presented a method to derive the ground-state magnetization 
curve in the thermodynamic limit from the finite-cluster calculation 
by the size scaling based on the conformal invariance.\cite{cft} 
The obtained curve of the $S=1$ antiferromagnetic chain 
successfully realized the experimental results 
of the magnetization measurements on NENP qualitatively. 
The method was also applied to get theoretical magnetization curves 
of some other one-dimensional spin systems.\cite{tonegawa,hagiwara} 
However, the critical behavior near $H_c$ cannot be investigated  
by this method, 
because it can yield too few points near $H_c$ 
to determine the critical exponent of the magnetization curve 
by the standard curve fitting. 

In general, except for the 
Kosterlitz-Thouless transition\cite{kt}, 
the magnetization $m$ near the critical field behaves like 
\begin{eqnarray}
\label{delta}
m \sim (H-H_c)^{1/{\delta}},
\end{eqnarray}
for the second-order phase transition. 
The critical exponent $\delta $ is an important quantity to 
determine the universality class of the phase transition 
which does not depend on any detailed properties of each system. 
For the $S=1$ antiferromagnetic chain 
the exponent was deduced as $\delta =2$ from some effective 
Hamiltonian theories.\cite{minoru,affleck} 
For the $S=1/2$ bond alternating chain 
the bosonization method gave the same result.\cite{chitra} 
$\delta =2$ was also derived from the fermionic excitation 
with the dispersion $k^2$ which was numerically verified 
to be a good picture for both systems.\cite{sakai2,sakai3} 
In addition 
the argument of the equivalence between the magnetization 
process of antiferromagnetic chains and  
some integrable models of 
the crystal-shape profile lead to the same conclusion.\cite{akutsu} 
In any theories giving $\delta =2$, 
however, 
the original spin Hamiltonians were mapped into 
other solvable models with some crucial approximations. 
Thus it would be important to estimate $\delta$ 
for the original Hamiltonian directly in some numerical ways,  
to test these effective theories and to investigate 
unknown systems. 

In this paper 
we propose a size scaling method to estimate the critical exponent 
$\delta $ of quantum spin chains using the result of the finite 
cluster calculation. 
In order to examine the validity of the method, 
we apply it to the $S=1$ antiferromagnetic chain and the $S=1/2$ 
bond alternating chains. 
In addition 
a recent topic on the magnetization plateau of the $S=3/2$ 
antiferromagnetic chain with anisotropy is investigated 
by the method. 

%
%

At first we consider the $S=1$ antiferromagnetic Heisenberg chain 
for the explanation of the method.  
The following argument is easily to be applied to more generalized 
models. 
To investigate the magnetization process we consider 
the Hamiltonian   
\begin{eqnarray}
\label{ham}
&{\cal H}&={\cal H}_0+{\cal H}_Z, \nonumber \\
&{\cal H}_0& = \sum _j^L {\bf S}_j \cdot {\bf S}_{j+1}, \\ 
&{\cal H}_Z& =-H\sum _j^L S_j^z, \nonumber
\end{eqnarray}
under the periodic boundary condition. 
We restrict us on even-site systems to avoid the frustration. 
Throughout we use the unit such that $g \mu _B =1$. 
For $L$-site systems, 
the lowest energy of ${\cal H}_0$ in the subspace where 
$\sum _j S_j^z=M$ 
(the macroscopic magnetization is $m=M/L$) is denoted as $E(L,M)$. 
We assume the asymptotic form of the size dependence of the energy as 
\begin{eqnarray}
\label{energy}
{1\over L}E(L,M)\sim \epsilon (m) +C(m) { 1\over {L^{\theta}}} 
\qquad (L\rightarrow \infty), 
\end{eqnarray}
where $\epsilon (m)$ is the bulk energy and the second term 
describes the leading size correction. 
We also assume that $C(m)$ is an analytic function of $m$. 
For gapless cases the conformal field theory 
predicted $\theta =2$.\cite{cft} 
Since the method works better 
for faster convergence of the size correction as shown later, 
we can also accept the exponential decay like $e^{-L/\xi}$ 
which is reasonably expected for the ground state of the spin gap systems 
instead of $1/L^{\theta}$. 
We neglect 
the $m$-dependence of $\theta$ because it gives only higher order 
corrections which does not change the main result. 
If the bulk system has the critical behavior described by the form 
(\ref{delta}), 
$m$-dependence of the energy $\epsilon (m)$ 
near $m=0$ should have the form 
\begin{eqnarray}
\label{em}
\epsilon (m) \sim \epsilon (0)+ H_c m +Am^{\delta +1}, 
\end{eqnarray}
where $A$ is a positive constant and we assume $\delta >0$. 
Now we put $M=$0, 1 and 2 into the form (\ref{energy}) and use 
(\ref{em}). 
If $L$ is sufficiently large, 
$C(m)$ can be expanded with respect to $m$ 
as $C(1/L)\sim C(0)+C'(0)1/L+1/2C''(0)1/L^2\cdots$ for $M=1$. 
Thus we get the forms
\begin{eqnarray}
\label{e0}
&E(L,0)\sim &L\epsilon (0) +C(0) {1\over {L^{\theta-1}}}, \nonumber \\
&E(L,1)\sim &L\epsilon (0) +H_c +A {1\over {L^{\delta}}} \nonumber \\
&           &+C(0){1\over{L^{\theta -1}}} +C'(0){1\over {L^{\theta}}}
 +{1\over 2}C''(0){1\over {L^{\theta +1}}} \cdots, \nonumber \\
&E(L,2)\sim &L\epsilon (0) +2H_c +2^{\delta +1} A {1\over {L^{\delta}}} 
\nonumber \\
&           &+C(0){1\over{L^{\theta -1}}} +2C'(0){1\over {L^{\theta}}}
 +2C''(0){1\over {L^{\theta +1}}} \cdots . \nonumber 
\end{eqnarray}
If we define the quantity 
\begin{eqnarray}
\label{f0}
f(L)\equiv E(L,2)+E(L,0)-2E(L,1), 
\end{eqnarray}
the asymptotic behavior of $f(L)$ becomes 
\begin{eqnarray}
\label{f}
f(L)\sim A(2^{\delta +1}-2){1\over {L^{\delta}}} +C''(0){1\over
{L^{\theta +1}}} \qquad (L \rightarrow \infty ). 
\end{eqnarray}
When the second term of (\ref{f}) converges faster than the first one,
the exponent $\delta$ can be estimated from the size dependence of 
$f(L)$. 
Thus the necessary condition under which the method gives the 
correct value of $\delta$ is 
\begin{eqnarray}
\label{condition}
\theta > \delta -1.
\end{eqnarray}
Therefore 
we have to check the condition 
that $E(L,0)/L$ converges faster than $1/L^{\delta -1}$ 
after determining $\delta$. 
Using the calculated values of $f(L)$ and $f(L+2)$, 
the exponent $\delta$ can be estimated by the form
\begin{eqnarray}
\label{estimate}
\ln \big( {{f(L)}\over{f(L+2)}} \big) /\ln \big({{L+2}\over L}) 
\sim \delta +O\big( {1\over {L^{\theta -\delta +1}}}\big). 
\end{eqnarray}
The convergence of the size correction is guaranteed by the condition 
(\ref{condition}). 
Thus the extrapolation of the $L$-dependent exponent 
$\delta (L,L+2)$ defined by the 
left hand side of (\ref{estimate}) 
gives an estimation of $\delta$. 

Note that the method can be easily generalized for the behavior 
around a finite magnetization $m_0$, which is described as 
$m-m_0 \sim (H-H_c)^{1/\delta}$. 
In this case we have only to change the form (\ref{f0}) into 
\begin{eqnarray}
\label{fm}
f(L)\equiv E(L,M_0+2)+E(L,M_0)-2E(L,M_0+1),
\end{eqnarray}
where $M_0=Lm_0$. 
In addition the method can be applied even to gapless cases 
where $H_c$ might be zero. 
In the following argument we don't mention the value of $H_c$ 
but we concentrate on the estimation of $\delta$. 

\begin{figure}[htb]
\begin{center}
\mbox{\psfig{figure=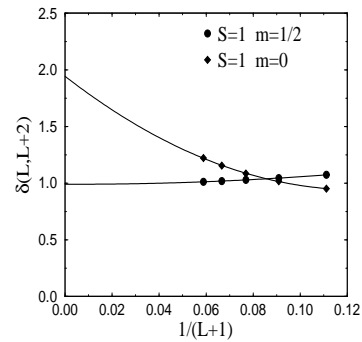,width=5cm,height=5cm,angle=-90}}
\end{center}
\caption{
$L$-dependent exponent $\delta (L,L+2)$ plotted versus $1/(L+1)$ 
for $m=1/2$ (solid circle) and $m=0$ (solid diamond) of the $S=1$ 
antiferromagnetic chain. 
The extrapolated results are $\delta =0.99 \pm 0.01$ and 
$\delta =1.9 \pm 0.1$, respectively. 
\label{fig1} }
\end{figure}

\begin{figure}[htb]
\begin{center}
\mbox{\psfig{figure=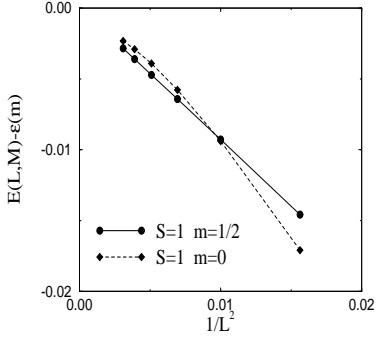,width=5cm,height=5cm,angle=-90}}
\end{center}
\caption{
Size correction of the ground state energy per site $E(L,M)-\epsilon
(m)$ of the $S=1$ antiferromagnetic chain plotted versus $1/L^2$ 
for $m=1/2$ (solid circle) and $m=0$ (solid diamond). 
It converges just like $1/L^2$ for $m=1/2$, while 
obviously faster than $1/L^2$.  
\label{fig2}
}
\end{figure}

%
%
For the behavior of the magnetization curve around $m=1/2$ 
of the $S=1$ antiferromagnetic chain, 
the $L$-dependent exponent $\delta (L,L+2)$ derived from  
the form (\ref{fm}) using 
the finite cluster results of $E(L,M)$ up to $L=18$ calculated 
by Lanczos algorithm 
is plotted versus $1/(L+1)$ in Fig. \ref{fig1}. 
Fitting the quadratic function 
$\delta (L,L+2) \sim \delta +a/(L+1)+b/(L+1)^2$ to the data, 
the extrapolated value is determined as $\delta =0.99 \pm 0.01$, 
based on the standard least-square method. 
The result leads to the conclusion $\delta =1$, 
which is reasonably expected for the gapless point $m=1/2$. 
To check the condition (\ref{condition}) 
$E(L,M)-\epsilon (m)$ for $m=1/2$ of the $S=1$ antiferromagnetic chain 
is plotted versus $1/L^2$ in Fig. \ref{fig2}, 
where the value of $\epsilon (m)$ was estimated by fitting of the quadratic
function of $1/L$.  
The plot suggests $\theta =2$ in the form (\ref{energy}) 
which is consistent with 
the prediction of the conformal field theory.\cite{cft} 
Then the condition (\ref{condition}) is satisfied. 

%
%

%
%
 
In Fig. \ref{fig1} we also show 
the plot of $\delta (L,L+2)$ based on the form (\ref{f0}) 
versus $1/(L+1)$ 
for $m=0$ of the $S=1$ antiferromagnetic chain 
which is more interesting because the system has the Haldane gap 
which vanishes at $H_c$. 
The extrapolated result is $\delta =1.9 \pm 0.1$ which suggests 
$\delta =2$, 
as predicted by the above effective Hamiltonian theories. 
The plot of $E(L,M)-\epsilon (m)$ versus $1/L^2$ for $m=0$ 
in Fig. \ref{fig2} obviously 
shows that the size correction in the form (\ref{energy}) decays 
faster than $1/L^2$ 
in contrast to the plot for $m=1/2$. 
It implies $\theta >2$ and the condition (\ref{condition}) 
which is $\theta > 1$ in this case 
is also satisfied for $m=0$. 

\begin{figure}[htb]
\begin{center}
\mbox{\psfig{figure=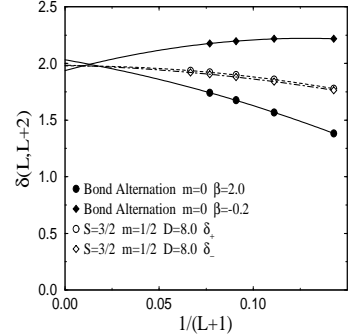,width=5cm,height=5cm,angle=-90}}
\end{center}
\caption{
$L$-dependent exponent $\delta (L,L+2)$ plotted versus $1/(L+1)$ 
for the $S=1/2$ 
bond alternating chain with $\beta =2.0$  
(solid circle) and $\beta=-0.2$ (solid diamond).  
The extrapolated results are $\delta =2.03 \pm 0.03$ and 
$\delta =1.9 \pm 0.1$, respectively. 
$L$-dependent exponents $\delta_{\pm} (L,L+2)$  
associated with the magnetization plateau at $m=1/2$ 
are also plotted versus $1/(L+1)$ 
for the $S=3/2$ antiferromagnetic chain with the single ion anisotropy
$D=8.0$; $\delta _+$: open circle and $\delta _-$: open diamond. 
The extrapolated results are $\delta_+ =1.98 \pm 0.04$ and
$\delta_- =1.99 \pm 0.04$, respectively.
\label{fig3}
}
\end{figure}

\begin{figure}[htb]
\begin{center}
\mbox{\psfig{figure=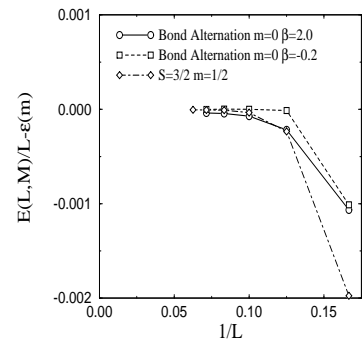,width=5cm,height=5cm,angle=-90}}
\end{center}
\caption{
Size correction of the ground state energy per site $E(L,M)-\epsilon
(m)$ of the $S=1$ antiferromagnetic chain plotted versus $1/L$ 
for the $S=1/2$ bond alternating chain ($m=0$) with $\beta=2.0$ 
(open circle) and $\beta=-0.2$ (open square), 
and the $S=3/2$ antiferromagnetic chain ($m=1/2$) with the single 
ion anisotropy $D=8.0$ (open diamond). 
It converges faster than $1/L$ in all the cases. 
We plot the original values times $10^4$ only for the bond alternating
chain with $\beta =-0.2$. 
\label{fig4}
}
\end{figure}

Next we investigate the $S=1/2$ bond alternating chain as another example 
with the spin gap between the singlet ground state  
and the triplet first excited state. 
The Hamiltonian is defined as 
\begin{eqnarray}
\label{hamb}
&{\cal H}&={\cal H}_0+{\cal H}_Z, \nonumber \\
&{\cal H}_0& = \sum _j^L {\bf S}_{2j-1} \cdot {\bf S}_{2j} 
-\beta  \sum _j^L {\bf S}_{2j} \cdot {\bf S}_{2j+1},\\ 
&{\cal H}_Z& =-H\sum _j^{2L} S_j^z, \nonumber
\end{eqnarray}
where $2L$ $S=1/2$ spins are included in the systems.  
The system has the gap except for $\beta =-1$ where it is 
the uniform $S=1/2$ antiferromagnetic chain. 
We chose two typical values of $\beta$; 
(i)$\beta =2.0$ and (ii)$\beta =-0.2$, 
which correspond to the ferromagnetic-antiferromagnetic 
and antiferromagnetic-antiferromagnetic alternating chains, respectively. 
In the latter case particularly the finite size effect is larger 
in the vicinity of the gapless point $\beta =-1$. 
Thus we study only for a smaller $\beta$ (=$-0.2$) than realistic cases.  
The universality argument, however, 
justifies that the critical exponents are independent of $\beta$ 
except for $\beta =-1$, 
because the system with $m=0$ is in a common phase for $\beta \not= -1$. 
In order to estimate the exponent $\delta$ around $m=0$ of the system 
(\ref{hamb}), 
the $L$-dependent exponent $\delta (L,L+2)$ 
up to $L=12$ is plotted versus $1/(L+1)$
for $\beta$=2.0 and $-0.2$ 
in Fig. \ref{fig3}. 
The same extrapolation as the $S=1$ chain results in 
$\delta =2.03 \pm 0.03$ and $1.9 \pm 0.1$ for 
$\beta =$2.0 and $-0.2$, respectively. 
The results are also consistent with $\delta =2$ 
predicted by some theories discussed above. 
We also have to check the condition (\ref{condition}) 
which is $\theta > 1$ because of $\delta =2$. 
In Fig. \ref{fig4} $E(L,M)-\epsilon (m)$ for $m=M=0$ of the system 
(\ref{hamb}) with $\beta =2.0$ and $-0.2$ is plotted versus $1/L$. 
It obviously shows a faster convergence of the size correction for 
the ground state energy per site than $1/L$, 
which implies that the condition is satisfied. 

%
%
%
%

Finally we apply the method to the $S=3/2$ antiferromagnetic chain 
with the single-ion anisotropy. 
The system is described by the Hamiltonian (\ref{ham}) 
with the anisotropy term $D\sum _j^L(S_j^z)^2$ 
added to ${\cal H}_0$. 
Recently an argument based on the analogy to the quantum Hall effect 
suggested that the ground state magnetization curve possibly had a plateau 
just at $m=1/2$ which corresponds to $1/3$ of the saturation moment 
and the singular part of the magnetization 
near the plateau was proportional 
to $\sqrt{|H-H_c|}$ where $H_c$ is the critical field at either 
edge of the plateau.\cite{oshikawa}  
The plateau was verified to exist for $D>D_c=0.93$ by finite cluster
analyses and size scaling techniques.\cite{sakai4} 
However, the form of the singularity near the edge of the plateau 
has not been derived by any numerical studies on the original 
Hamiltonian. 
Thus this problem is one of interesting examples to investigate 
by the method presented in this paper. 
We consider a sufficiently large $D$ so that the magnetization 
curve has a clear plateau at $m=1/2$. 
The two critical fields $H_{c\pm}$ are denoted such that 
the curve has a plateau for $H_{c-} <H< H_{c+}$.  
They can be given by 
\begin{eqnarray}
\label{lim}
E(L,{L\over 2}\pm 1)-E(L,{L\over 2}) \rightarrow \pm H_{c\pm} 
\quad (L\rightarrow \infty), 
\end{eqnarray}
although we don't consider the value of $H_{c\pm}$ here. 
To investigate the singularity of the magnetization curve, 
the critical exponents $\delta _{\pm}$ are defined as 
\begin{eqnarray}
\label{delta2}
m-{1\over 2} \sim (H-H_{c+})^{1/{\delta _+}}, \\
{1\over 2}-m \sim (H_{c-}-H)^{1/{\delta _-}}.
\end{eqnarray}
To estimate $\delta _{\pm}$ 
we have only to change $f(L)$ into $f_{\pm}(L)$ defined as 
$f_{\pm}(L)\equiv \pm[E(L,{L\over 2}\pm 2)+E(L,{L\over 2}) 
-2E(L,{L\over 2}\pm 1)]$ 
and extrapolate the $L$-dependent exponents $\delta _{\pm}(L,L+2)$ 
defined by the left hand side of the equation (\ref{estimate}) 
using $f_{\pm}(L)$ instead of $f(L)$. 
In Fig. \ref{fig3} we show 
the plot of $\delta _{\pm}(L,L+2)$ versus $1/(L+1)$ up to $L=14$ 
for $D=8.0$. 
The extrapolated results are $\delta _+=1.98 \pm 0.04$ and 
$\delta _-=1.99 \pm 0.04$,
which imply $\delta _+=\delta _-=2$ as suggested by the analogy 
to the quantum Hall effect. 
We also check the condition (\ref{condition}) by the plot of 
$E(L,M)-\epsilon (m)$ versus $1/L$ for $M=L/2$ and $m=1/2$ in 
Fig. \ref{fig4} which suggests that the size correction decays 
faster than $1/L$. 
To avoid large finite size effects we considered only a large value of $D$ 
($=8.0$) which is not realistic. 
But it is expected the result $\delta =2$ is always true for
$D>D_c(=0.93)$ because the transition at the critical field belongs 
to a common universality class. 

Recently the magnetization plateau was also investigated 
on the $S=1/2$ bond alternating chain with the next-nearest 
neighbor interaction\cite{totsuka} by the bosonization technique 
which lead to $\delta =2$ at the edge of the plateau at $m=1/4$. 
The result suggests the transition belongs to the same universality 
class as that of the anisotropic $S=3/2$ chain. 
 
In summary 
a finite size scaling method 
to estimate the critical exponent $\delta$ associated with 
the magnetization curve around the critical magnetic field corresponding
to the amplitude of the spin gap of quantum spin chains was proposed 
and applied to the $S=1$ antiferromagnetic chain and $S=1/2$ bond 
alternating chain. 
In addition the behavior of the magnetization curve around the edges 
of the plateau of the anisotropic $S=3/2$ antiferromagnetic chain was 
investigated by the method. 
All the results indicated the same conclusion $\delta =2$.  

%
%
We would like to thank Dr. K. Totsuka for sending his preprint 
and interesting discussions. 
We also thank 
the Supercomputer Center, Institute for
Solid State Physics, University of Tokyo for the facilities
and the use of the Fujitsu VPP500.
This research was supported in part by Grant-in-Aid 
for the Scientific Research Fund from the Ministry 
of Education, Science, Sports and Culture (08640445).

\end{document}